\documentclass[useAMS,usenatbib, onecolumn]{mn2e}

\newcommand{\be}{\begin{equation}}
\newcommand{\bs}{\begin{subequations}}
\newcommand{\cm}{{~\:\rm cm}}

\newcommand{\cc}{{~\:\rm cm^{-3}}}
\newcommand{\dpar}[2]{\frac{\partial{#1}}{\partial{#2}}}
\renewcommand{\div}{{\nabla \cdot}}
\renewcommand{\d}[2]{\frac{d {#1}}{d {#2}}}
\newcommand{\dd}[2]{\frac{d^2 {#1}}{d {#2}^2}}
\newcommand{\eexp}[1]{{\rm e}^{#1}}
\newcommand{\ee}{\end{equation}}
\newcommand{\es}{\end{subequations}}

\newcommand{\ergs}{{{~\:\rm erg}\; \rm s^{-1}}}
\newcommand{\g}{{~\:\rm g}}

\newcommand{\keV}{{~\:\rm keV}}

\newcommand{\kh}{{\bm k_h}}
\newcommand{\kpc}{{~\:\rm kpc}}
\newcommand{\s}{{~\:\rm s}}
\renewcommand{\v}{{\bm v}}
\newcommand{\yr}{{~\:\rm yr}}
\newcommand{\x}{{\bm x}}
\newcommand{\xx}{{\bm \xi}}

\usepackage{amsmath}                
\usepackage{amsfonts}               
\usepackage{amssymb}                
\usepackage[british,english]{babel} 
\usepackage{bm}                     
\usepackage{color}                  
\usepackage{dcolumn}                
\usepackage{graphicx}               

\begin{document}


\title[Rayleigh--Taylor Instability]{On the Rayleigh--Taylor Instability 
of  Radio Bubbles  in Galaxy Clusters}


\author[F. Pizzolato and N. Soker]
{Fabio Pizzolato$^1$
\thanks{E-mail fabio@physics.technion.ac.il}
and
Noam Soker$^1$
\\
$^1$Department of Physics, Technion--Israel Institute of Technology, 
Haifa 32000 Israel
}

\date{\today}

\pagerange{\pageref{firstpage}--\pageref{lastpage}} \pubyear{2006}


\maketitle

\label{firstpage}


\begin{abstract}

We consider the Rayleigh--Taylor instability in the early evolution
of the rarefied radio bubbles (cavities) observed in many cooling--flow 
clusters of galaxies.
The top of a bubble becomes prone to the Rayleigh--Taylor instability as 
the bubble rises through the intra--cluster medium (ICM).
We show that while the jet is  powering the inflation,
the deceleration of the bubble--ICM interface  is able to reverse the 
Rayleigh--Taylor instability criterion.
In addition,   the inflation introduces a drag effect which increases 
substantially the instability growth time.  The combined action of
these two effects considerably delays the onset of the instability.
Later on, when the magnitude of the deceleration  drops or the jet fades, the 
Rayleigh--Taylor and the Kelvin--Helmholtz instabilities set 
in and eventually disrupt the bubble.
We conclude that the initial deceleration and drag, albeit unable to
prevent the disruption of a bubble,  may significantly lengthen its
lifetime, removing the need to  invoke stabilising magnetic fields.

\end{abstract}




\section{Introduction}
\label{s-intro}

The recent high spatial--resolution observations show that many
clusters and groups of galaxies harbour bubbles (cavities) devoid of 
X--ray emission.
Examples include 
Perseus  \citep{Fab00},
Hydra~A  \citep{McN00},
Abell~2052 \citep{Bla01},
Abell~2597 \citep{McN01},
MKW~3s \citep{Maz02},
HCG~62 \citep{Vrt02},
Abell~4059 \citep{Hei02}
Abell~478 \citep{Sun03}, and
MS~0735.6+7421 \citep{McN05};
see also  \citet{Bir04}  for a systematic study.
These bubbles are very low density  regions inflated by the jets
launched by the active galactic nuclei (AGN) sitting at the
centres of these clusters.
The lack  of strong shocks at the rims of these bubbles shows that their
inflation is quite  gentle, and the bubbles are in approximate pressure
equilibrium with the surrounding environment.
On account of their lower density with respect to the
outer environment, the bubbles rise buoyantly and sweep up some of the
intra--cluster medium (ICM), as evident by the enhanced emission at their 
rims.

It would seem that  the top of these bubbles is prone to 
the  Rayleigh--Taylor Instability (hereafter, RTI). Indeed,
the gravitational acceleration of the cluster pushes downwards, and
the dense swept-up gas shell  lies on the top of the thinner one inside the
bubble itself.
This RTI would be able to tear the bubble apart in few characteristic
e--folding times
\be
\label{e-rt}
t_{\rm RT} = \left(\frac{\lambda_h}{2\:\pi\: |g|}\right)^{1/2} =
2.2  \times 10^6   \yr \;
\left(\frac{\lambda_h}{\kpc}\right)^{1/2} \;
\left(\frac{|g|}{10^{-7}\cm\s^{-2}}\right)^{-1/2},
\ee
where $|g|$ the modulus of the cluster gravitational acceleration and
$\lambda_h$ is the wavelength of the RT perturbation.
The characteristic time~\eqref{e-rt} is about one order of
magnitude less than the estimated ages of most bubbles \citep{Bir04}.
This observation has prompted many authors 
\citep[e.g.][]{Bru01, Kai05, Rey05, Jon05}
to invoke an ordered  magnetic  field at the edge of the bubble to
stabilise the bubble against the RTI.
This scenario, however, may lead to some difficulties:
if the magnetic field is ordered  on a scale of the same order as
the size of the bubble, its effect should be highly non--isotropic:
it should be able to suppress very efficiently the 
RTI modes along the magnetic field lines, but not perpendicular to them.
This is because the magnetic field cannot exert stress countering the 
instability a direction perpendicular to the field lines.
If the field is tangled on scales much smaller than the bubble size, then
it will have no effect on the large-wavelength modes which are those
that tear the bubble apart.

In this paper we argue that a magnetic field is actually not
necessary to explain the long lives of the radio bubbles.
Extending the work by~\citet{Sok02b} we shall show that
during the first stages of its inflation a  bubble  is  stable.
In  the early phase the inflation of the bubble is powered by 
a jet launched by the central AGN. While the jet is active 
the bubble radius grows with time as  $R_b(t) \propto t^{3/5}$,
implying that the acceleration $\ddot R_b$ is {\em negative}.
On  the top of the bubble, then, this deceleration
overcomes gravity, and the configuration of the bubble is
RT--stable.
Besides, the inflation introduces a drag effect which greatly 
reduces the instabilities growth rate.
Only later, when $|\ddot R_b|$ decreases and the 
gravitational acceleration prevails, or the jet fades,  
does the instability set in, tearing the bubble apart.
We show that the delay of the RTI  onset is more than enough to
explain the survival of old bubbles. 

This  paper is laid out as follows; in \S~\ref{s-bubbles} we discuss the
properties of the bubbles, and sketch the physics developed in  this
paper. In \S~\ref{s-equations} we derive the equations of the
RTI in a  bubble actively inflated by a jet.
In \S~\ref{s-numerical} 
these equations  are solved numerically,  and the results are  discussed
in \S~\ref{s-discussion}. 
The last  stage of  the life of a bubble, after the extinction of the 
jet,   is addressed in \S~\ref{s-last}. We summarise in \S~\ref{s-summary}.
Readers not interested in the mathematical details may harmlessly skip
\S~\ref{s-equations} and~\S~\ref{s-numerical}, jumping directly  from  
\S~\ref{s-bubbles} to \S~\ref{s-discussion}.

\section{The Rayleigh--Taylor Instability in Bubbles}
\label{s-bubbles}

In this Section we refine the treatment by \citet{Sok02b} of the RTI
of a bubble inflated by an AGN jet.
The jet stops at a distance $r_c$ from the cluster core, and hence it
becomes the expansion centre  of the  bubble.
The injected jet's material passes  through a strong shock and forms
a very hot bubble, which later on  keeps in approximate pressure
equilibrium with its surroundings.
On account of its low density, the bubble rises buoyantly into the ICM;
during this process the expanding bubble sweeps up the ICM to 
form a thick shell of denser gas ahead of itself, 
as evident from  the local enhanced X--ray luminosity. 
The X--ray observations also show a  lack
of strong shocks ahead of the bubble, so the swept--up shell and the gas
inside the bubble move about with the same velocity.
Because the shell density $\rho_s$ is much higher than the density $\rho_b$
inside the bubble,
 and since the gravitational acceleration is directed downward, the
flow at the bubble's top would be RT unstable if only gravity is 
considered.
This simple picture, however, overlooks the bubble's inflation process.
We therefore consider a local frame of reference attached to a point $P$ on
the surface of the expanding bubble, and moving with it.
A test particle  in the non--inertial frame 
attached to  $P$ feels the radial (i.e., normal to the bubble's surface)
acceleration
\be
\label{e-accel}
\tilde g = - |g| \; \cos\theta - \ddot{R}_b(t),
\ee
where $g$ is  the cluster's gravitational acceleration,    
$\theta$  the angle between the orthogonal to the bubble 
surface in $P$ and the radial direction  of  the cluster, 
and $R_b(t)$ is the bubble radius at the instant $t$.
If the jet supplies energy at the constant rate of $\dot E$, then
\be
\label{e-rtime}
R_b(t) = \alpha \;  \left({\dot E} \:  t^3/\rho_a \right)^{1/5}
\ee
\citep{Cas75,Dok02}, where $\rho_a$ is the density of the ambient ICM.
The dimensionless parameter $\alpha$ depends on the equation of
state  (EOS) of the  gas inside the bubble:
for a non--relativistic EOS $\alpha=0.929$, while for a relativistic EOS
$\alpha = 0.793$ \citep{Dok02}.
For bubbles in cooling flow clusters the EOS is most likely to be
somewhere in between these two extremes.
It is worth noticing that Equation~\eqref{e-rtime} assumes a 
uniform  ambient density, which is not true for the stratified atmosphere
of a cluster the bubble is embedded in. Nevertheless,  in 
the following we shall adopt  Equation~\eqref{e-rtime} since it
is but a secondary source of error. Moreover, if the cluster's 
negative density gradient is considered, then the bubble's top has the 
largest acceleration, making it more stable. Since the bubble's top front 
is the least stable region, considering the density gradient will make the 
stabilising effect studied here more favourable.

The  normal  acceleration of a test particle residing on the 
bubble--ICM boundary  in the non--inertial frame is therefore
\be
\label{e-geff}
\tilde g = - |g| \; \cos\theta + \tfrac{6}{25} \; \frac{R_b(t)}{t^2}.
\ee
Near  the  bottom of the bubble $\theta\sim\pi$, so  $\tilde g$ is
always positive, meaning that the cluster's gravitational acceleration
and $\ddot R_b$ have the same direction, and  the total acceleration
pulls towards  the cluster centre.  In this segment, the bubble is
RT stable, because the lighter plasma (inside the bubble)
lies above the denser one.
Near the top of the bubble ($\theta\sim 0$)  the 
acceleration~\eqref{e-geff} may be either positive or negative.
At early stages ($t$~small), $\ddot R_b$ is large enough to
overcome gravity, thus making  $\tilde g$ positive.
The acceleration pulls upwards, and on account of the density
stratification  even this  segment is RT--stable,
without the need of any magnetic field \citep[as pointed out by][]{Sok02b}.
On top of this, as discussed in \S~\ref{s-equations}, the inflation
introduces a drag effect on the perturbation, akin to the Hubble drag
affecting the perturbations growth  in  an expanding Universe.
Both in the cosmological context and here, this drag reduces the 
perturbations growth rate, further delaying the RTI onset.
Later on, however  the  gravitational pull 
will prevail, and the  RTI commence.
The delay introduced by the early contribution of $\ddot R_b$ 
and by the drag term is more than enough 
to lengthen the life time of the bubble to a significant amount.
The mathematical details justifying this statement are
worked out in the following Section.
Readers not particularly eager to afford them may 
skip~\S~\ref{s-equations} and~\S~\ref{s-numerical}  
and go directly to Section~\ref{s-discussion}.

\section{Rayleigh--Taylor Instability of an Expanding Bubble}
\label{s-equations}

In this Section  we derive the equations for  the Rayleigh--Taylor
instability of an expanding bubble.
As customary  in any  perturbation analysis,
it is convenient to decouple the  ``background'' motion of the bubble
(i.e., its expansion and  uprising) from the motion associated to
the true instability.
To this aim we introduce a frame of reference moving with the
expanding bubble. We consider an inertial frame of reference $K$, the origin
of which coincides with the cluster's centre.
In this frame of reference the bubble's centre $O$ has coordinate 
$\bm  r_b(t)$.
We define a second frame  $K''$  with its axes parallel to those of $K$,
and  centred on $\bm r_b$.
Finally, we define the non--inertial frame $K'$ with the same centre as $K''$,
which co--moves with the
expanding bubble: in other words, the coordinates of any point
expanding with the bubble are constant in the frame $K'$.
The relation between the coordinates $\bm x''$ and $\bm r'$,
referred to the frames $K''$ and $K'$ respectively,  is
\be
\bm x'' = a(t) \; \bm r',
\ee
where (from Equation~\eqref{e-rtime}) we define
the expansion scale factor
\be
\label{e-scale}
a(t) =  R_b(t)/R_b(t_0) = \left(t/t_0\right)^{3/5},
\ee
where $t_0$ is an arbitrary reference time.
Putting together all the previous equations, the relation between the
inertial  frame $K$ and  the co--moving frame $K'$ is
\be
\label{e-k2k1}
\bm x = \bm r_b(t) + a(t) \; \bm r'.
\ee
This formula allows us to write down  a set of hydrodynamic equations in the
co--moving frame $K'$. The advantage of this approach is that the 
unperturbed
equations in $K'$  are time--independent, allowing a more direct
approach to the subsequent perturbation analysis.

We write  the usual gas dynamic equations in the inertial frame $K$ 
\citep[e.g.][]{Lan87}
\bs
\label{e-gasd}
\begin{align}
& \dpar{\rho}{t}  + \div(\rho \; \v) = 0
\\
&  \rho \: \frac{d \v }{d t}  + \nabla P - \rho \; \bm g = 0.
\end{align}
\es
From  the coordinate transformation~\eqref{e-k2k1} and the equality $t = t'$ 
($K$ and $K'$ have the same time units),
we retrieve the relations between the derivatives in $K$ and $K'$:
\bs
\begin{align}
&\dpar{}{t} = \dpar{}{t'} - \frac{\bm v_b +  \dot{a} \; \bm r'}{a}  \; 
\cdot\nabla'
\\
&\nabla = \frac{1}{a} \; \nabla'.
\end{align}
\es
In the first equality we assume that during the relevant
evolutionary phases for our study the buoyancy velocity
$\bm v_b =  \dot{\bm r}_b$ is negligible in comparison to
the expansion velocity $\dot{a} \; \bm r'$.
Plugging these into Equations~\eqref{e-gasd},
after some algebra, we get the following set of equations in the co--moving
frame $K'$:
\bs
\label{e-moving}
\begin{align}
\label{e-movingcon}
& \dpar{\rho'}{t'}  + \div'(\rho' \; \v') = 0
\\
\label{e-movingeul}
&  \rho' \:
\left(\frac{d' \v' }{d t'} +  \frac{\ddot{a}}{a}\: \bm  r' + 2 
\:\frac{\dot{a}}{a}\: \bm v' \right) +
\frac{1}{a} \;
\biggl( \nabla' P' - \rho' \;\bm g\biggr) = 0,
\end{align}
\es
where we have introduced the new co--moving variables
\bs
\begin{align}
& \rho' = \rho \; a^3
\\
&\v'  =  \frac{\bm v - \dot{a}\; \x'}{a}
\\
&P' = P\; a^2,
\end{align}
\es
and $d'/d t' = \partial_{t'} + \bm v'\cdot\nabla'$ is the convective
derivative in the co-moving frame.

\subsection{The Unperturbed Configuration}
\label{s-unperturbed}

As noted in the discussion  following Equation~\eqref{e-geff},
the most prone segment  to the RTI is the top of the bubble.
Where necessary, therefore, we shall limit our analysis to this portion of 
the bubble.
Before studying the perturbations, it is necessary to define a
``background'' unperturbed configuration.
In the present case it is, obviously, the configuration of an
expanding bubble. This is described by the set of equations~\eqref{e-moving}
with $\bm v'=0$.
The co--moving continuity equation~\eqref{e-movingcon} yields
$\partial_{t'} \rho'=0$, i.e. $\rho'$ is constant with time;
the density $\rho$ in the inertial frame is
$\rho\propto a^{-3}$, i.e.
the mass contained within any given volume expanding with   the bubble
is constant, as expected.

As our X--ray observations show, there is no signature of strong
shocks between the bubble interior and the overlying swept--up shell.
For this reason, we assume that the shell and the  adjacent segment of
the bubble interior move with the same speed, and the only difference
between them is in density.
We denote with  $R$ the co--moving coordinate  of the interface between
the thick shell and the bubble interior.
The unperturbed density profile undergoes a sharp discontinuity across the
edge of the bubble, passing from  the low value $\rho_b$ inside the bubble
($r'<R$) to the ambient value $\rho_a$ outside it ($r'>R$).
We put $\rho_s\sim \rho_a$, i.e. we neglect the density enhancement
ahead of the bubble  due to the swept--up material,
since its  effect is small.
Our unperturbed co--moving density profile is then
\be
\label{e-dprof}
\rho'\: =
  \begin{cases}
  \rho_a       & \qquad r' > R
  \\
  \rho_b       & \qquad r' < R.
  \end{cases}
\ee
The co--moving  unperturbed static Euler  equation
\be
\nabla' P' = \rho' \left(\bm g - \ddot{a}\; \bm r' \right),
\ee
can be projected in the radial  and tangent direction with respect to
the co--moving frame $K'$:
\begin{align}
\label{e-equil}
&\dpar{P'}{r'}  = \rho' \;(g_{r'} - \ddot{a}\; r'),
\\
&\nabla_h P' = \rho' \; g_h,
\end{align}
where $g_{r'}$ and $g_h$ are respectively the projections of the
cluster  gravitational acceleration parallel and perpendicular to
the direction of the bubble's radius.
It is quite apparent that the term $g_h$ is
non--zero if the point $P$ is not aligned with the line joining the
bubble and the cluster centres.  On the top of the bubble, which
concerns us most, $g_h\sim 0$, so $\nabla_h P\sim 0$, and
the pressure gradient is purely radial.
Here, Equation~\eqref{e-equil} defines a local hydrostatic equilibrium
in the  effective  gravity
\be
\label{e-geff2}
g_{r'} -  \ddot{a}\; r' = - \bigl[ |g| + \ddot{a}(t)\; r'\bigr] = - 
\tilde{g}.
\ee
The overall effective gravity is {\em smaller} than expected from the
cluster's contribution alone, since $\ddot a(t) <0$.

\subsection{Lagrangian Perturbation Analysis}
\label{s-lagrange}

Having  determined the properties of the equilibrium solution,
we pass to study the perturbation to the inflating bubble.
For the sake of simplicity, from now on  we omit all the primes 
referred to the
co--moving frame $K'$. It should be clear, however, that all the
variables hereafter refer to that frame.

The perturbation analysis is most easily performed in the Lagrangian
formalism  (see e.g. \citealp{Sha83} for a nice introduction).
Let  $\xx(\x, t)$ be  the displacement of a fluid element due to the
perturbation. The  Lagrangian density perturbation $\Delta \rho$,
i.e. the perturbation  measured in a frame of
reference moving with the unperturbed flow,   is
\be
\label{e-pcont}
\Delta \rho  = - \rho \; \div \xx.
\ee
In the context of the study of the RTI, it is customary to
assume that the perturbation is incompressible. This assumption
holds if the  RTI growth time (Equation~\eqref{e-rt})
is shorter than the time
\be
\label{e-sound}
t_{\rm sound} \sim \lambda/c_s
\sim
1.0 \times 10^6 \yr \;
\left(\frac{\lambda_h}{\kpc}\right) \;
\left(\frac{T}{10\keV}\right)^{-1/2}.
\ee
taken by a sound  wave to  cross the perturbation itself.
The rationale of this request is that if $t_{\rm sound}$ is very short,
then the sound waves are very effective in keeping the pressure equilibrium
within the portions of the  perturbation.
In our case the incompressibility requirement $t_{\rm RT} \ll t_{\rm sound}$
does not seem to hold.
However,  the compressibility affects little
the growth rate of the RT perturbation
when  the density ratio $\rho_a/\rho_b$ is large, as occurs in
the present case \citep[see e.g][]{Bak83, Liv04}.
Therefore, it is safe to  adopt the incompressibility
assumption $\Delta \rho = 0$: from
Equation~\eqref{e-pcont}, this  is equivalent to
\be
\label{e-unc}
\div \xx = 0.
\ee
The $a$-th component of the perturbed Euler equation reads
\be
\label{e-eul1}
\rho \left[\dd{\xi_a}{t} + 2\:\frac{\dot{a}}{a}\: \d{\xi_a}{t} \right]
+ \nabla_a(\Delta P) - (\nabla_s P_0) \; \nabla_a \xi_s \; = 0,
\ee
where the pressure gradient may be expressed in terms of the
unperturbed Euler equation $\nabla P = \rho \: \tilde{\bm g}$.
It is convenient to
decompose the displacement vector  $\xx$ to a radial component
$\xi$ and a tangential (or horizontal) component $\xx_h$:
\be
\xx  =
\left[ \begin{array}{c}
    \xi  \\
    \xx_h
    \end{array}
\right].
\ee
The condition of incompressibility~\eqref{e-unc}  then reads
\be
\frac{1}{r^2} \; \dpar{}{r}\bigl( r^2\;\xi \bigr) + \nabla_h \cdot \xx_h = 
0.
\ee
If we may neglect the radius of curvature of the bubble, this simplifies to
\be
\dpar{\xi}{r}  + \nabla_h \cdot \xx_h \sim 0.
\ee
This approximation is permitted so long as the size $\lambda$ of a perturbation
(which is stretched by the bubble  inflation itself)
is much smaller than bubble radius $R_b$.
Although we are also interested in perturbation with $\lambda \sim R_b$,
we will use this approximation throughout the rest of the paper, as it is 
adequate
enough for our goal of showing that no stabilising magnetic fields are needed
to explain the survival of old bubbles.

We take advantage of the fact that the unperturbed quantities
only depend on the radial  coordinate $r$ to
separate out the dependence  on  the horizontal coordinates:
\bs
\begin{align}
\xi       &= \xi(r, t)\; \eexp{i \kh \cdot \x}
\\
\xx_h     &= \xx_h(r, t) \;\eexp{i \kh \cdot \x}
\\
\Delta P  &= \Delta P(r, t) \; \eexp{ i \kh \cdot \x},
\end{align}
\es
where $\kh$ is the horizontal (i.e., tangential)   
co--moving wavenumber  vector.
The perturbed Euler equation~\eqref{e-eul1} reads, component by component,
\be
\rho \; \dd{}{t}
    \left[ \begin{array}{c}
    \xi \\
    \xx_h
    \end{array}
\right]
+
2\; \frac{\dot{a}}{a} \; \rho \; \d{}{t}
    \left[ \begin{array}{c}
    \xi \\
    \xx_h
    \end{array}
\right]
+ \frac{1}{a}\;
   \left[ \begin{array}{c}
    \partial{\Delta P}/\partial{r} \\
    i \; \kh\;  \Delta P
    \end{array}
\right]
+
\frac{\rho \;\tilde{g}}{a} \;
    \left[ \begin{array}{c}
    \partial \xi /\partial r  \\
    i\; \kh\;  \xi
    \end{array}
\right]
= 0,
\ee
where $\tilde g$ is defined by Equation~\eqref{e-geff2}.
We also have the (approximated)  condition of incompressibility
\be
\dpar{\xi}{r} + i\: \kh \cdot \xx_h = 0.
\ee
Dotting the horizontal component of the Euler equation  by $\kh$ and
eliminating from there $\xx_h \cdot \kh$ via the incompressibility
condition, we retrieve
\be
\Delta P =   - \rho \;
\left\{
\tilde{g} \;\xi \;+\;
\frac{a}{k_h^2} \; \left[\dd{}{t}\dpar{\xi}{r} + 2 \; \frac{\dot{a}}{a}\; 
\d{}{t} \dpar{\xi}{r} \right] \right\}.
\ee
Inserting  this into the radial  component, after some 
algebra we find
\be
\label{e-dyn}
\rho \; k_h^2\; \left[ \dd{\xi}{t} + 2\;\frac{\dot{a}}{a}\;
\d{\xi}{t} \right]
-
\dpar{}{r}\left[ \rho \left(\dd{}{t}\dpar{\xi}{r} \;+\; 2 \; 
\frac{\dot{a}}{a} \d{}{t} \dpar{\xi}{r}  \right)\right]
-
(\rho_a - \rho_b)  \; \delta(r - R) \;
\frac{\tilde{g}\;k_h^2}{a} \; \xi  = 0,
\ee
where we have neglected the spatial variation of the effective gravity
$\tilde{g}$  on the scales we are concerned with; the delta term stems from
the radial derivative of the unperturbed discontinuous density
profile~\eqref{e-dprof}.

We further simplify the RTI criterion given by  Equation~\eqref{e-dyn}.
In our frame of reference
the background is static, so $d / d t = \partial / \partial t$.
Equation~\eqref{e-dyn} must be solved separately inside and outside the
bubble, i.e. for $r\gtrless R$. The two branches are not independent
of each  other, but are linked by the  jump condition imposed by the delta
function in  Equation~\eqref{e-dyn}.
For $r\neq R$ Equation~\eqref{e-dyn} reduces to
\be
\label{e-updown}
k_h^2\;
\left( \partial^2_{t t}\xi  + 2\;\frac{\dot{a}}{a}\; \partial_t \:\xi 
\right)
-
\left(\partial^2_{t t} \:\partial^2_{r r} \xi + 2 \; \frac{\dot{a}}{a}\; 
\partial_t \: \partial^2_{r r} \: \xi   \right) = 0.
\ee
For this equation we look for solutions in the separate form
\be
\label{e-lp}
\xi(r, t) = \phi(t) \; \zeta(r).
\ee
We  find
\be
\dd{\zeta}{r} - k_h^2\; \zeta = 0,
\ee
so Equation~\eqref{e-updown}  has the piecewise solution
\be
\label{e-sol}
\xi(r, t)  =
\begin{cases}
\phi(t) \; \eexp{- k_h\: (r - R) } & \qquad r > R
\\
\phi(t) \; \eexp{k_h\: ( r - R)  } & \qquad r < R.
\end{cases}
\ee
The function $\phi(t)$ is determined by the jump condition across the
surface of discontinuity $r=R$. We  integrate
Equation~\eqref{e-dyn} over the  thin layer
$[R-\epsilon, R+\epsilon]$, where  $\epsilon\to 0$.
After some algebra we derive
\be
\label{e-jump}
\lim_{\epsilon \to 0} \:
\left[ \rho_0 \left(\dd{}{t}\dpar{\xi}{r} \;+\; 2 \; \frac{\dot{a}}{a} 
\d{}{t} \dpar{\xi}{r}  \right)\right]_{R-\epsilon}^{R+\epsilon} =
- \frac{\tilde{g}(R) \;k_h^2}{a} \; \phi(t)  \; (\rho_a - \rho_b),
\ee
where on the right-hand side  we have used the identity
$\phi(t)= \xi(r=R, t)$.

We now may insert the solution~\eqref{e-sol} into the jump
condition~\eqref{e-jump} to get our  equation for  the
evolution of the Lagrangian perturbation:
\be
\label{e-pert}
\dd{\phi}{t} + 2\; \frac{\dot{a}}{a} \d{\phi}{t} -
\frac{\tilde{g}\: k_h}{a} \; \frac{\rho_a - \rho_b}{\rho_a + \rho_b} \; \phi 
= 0,
\ee
where  the effective acceleration $\tilde{g}$ must be evaluated
on the top of the bubble.

Note that if  ${a}=$~constant this formula  would reduce to the familiar 
RTI equation
\be
\label{e-noinflation}
\dd{\phi}{t} = |g| \: k_h \; \frac{\rho_a - \rho_b}{\rho_a + \rho_b} \; 
\phi,
\ee
the growing solution of which is the  exponential 
$\phi \propto \eexp{n\; t}$, with
\be
n^2 = |g| \: k_h \; \frac{\rho_a - \rho_b}{\rho_a + \rho_b}.
\ee
If $\rho_a > \rho_b$, i.e. the denser fluid lies on top of the lighter one,
then $n^2>0$: the perturbations grows exponentially as in the
standard scenario of the  RTI.

In the case of the inflating bubble, it is convenient to
express  the coefficients of Equation~\eqref{e-pert} terms of
physical (rather than co--moving) variables. The effective gravity
on the top of the bubble  is
\be
\tilde{g} = |g(r_b + R_b)|  + \ddot{R}_b(t),
\ee
The wavelength  $\lambda_h=2\,\pi/k_h$
in all our equations is a  {\em co--moving} quantity,
so its magnitude  does not have a precise meaning
{\em per se}, but only when it forms a dimensionless quantity with
the curvature radius $R_b$ of the bubble.
The  dimensionless quantity   
\be
\varkappa = k_h(t)  \:R_b(t), 
\ee 
i.e. the ratio between the bubble circumference and the perturbation's
{\em physical} wavelength  is  frozen in time, 
since the perturbation is
stretched by the expansion of the bubble. In the last equation we have
written $k_h(t)$ to remark that here this wave number refers to a
wave in physical (i.e. non--comoving) coordinates. 
In order to be consistent with our planar approximation, the
factor  $\varkappa$  must be large.
Finally, since the bubble's density is always much lower than the 
ambient density 
\be
\label{e-lowd}
\rho_b \ll \rho_a,
\ee
we write down  the final
expression of the perturbation equation in the form
\be
\label{e-working}
\dd{\phi}{t} + \tfrac{6}{5}\; \frac{1}{t} \d{\phi}{t} -
\varkappa  \; \frac{|g|  + \ddot{R}_b(t)}{R_b(t)}\;\phi  = 0,
\ee
where we used Equation~\eqref{e-scale} to substitute for $\dot a/a$.
It is worth to remember that $\phi$ is a co--moving  variable:
its link with the  ``physical'' amplitude  $\phi_{\rm phys}(t)$,
i.e. the amplitude expressed in non--comoving coordinates is given by
the relation
\be
\label{e-phidef}
\phi(t) =  \phi_{\rm phys}(t) /  R_b(t).
\ee
The qualitative properties  of Equation~\eqref{e-working}
are briefly discussed  in the next Subsection.

\subsection{Properties of the Rayleigh--Taylor Instability of an Inflating  Bubble}
\label{s-properties}

Equation~\eqref{e-working} differs from the standard equation for
the Rayleigh--Taylor instability in three respects.

\begin{enumerate}

\item The term proportional to $d\phi/dt$. This is essentially due to
the fact that the unperturbed bubble  is inflating; this term is
therefore akin to the ``Hubble drag'' in the equations
for the evolution of cosmological perturbations in an expanding Universe
(see e.g. \citealp[page~470]{Pea99}).

\item The total acceleration $\tilde g$ is made up by two
contributions, the gravitational acceleration and the inflation 
deceleration.
We have already illustrated how this may affect the onset of the RTI
(see the discussion of Equation~\eqref{e-geff} in \S~\ref{s-bubbles}).

\item The last term in Equation~\eqref{e-working}  is
proportional to $R_b^{-1}$.
Again, this is an effect of the co--moving frame:
the factor $R_b^{-1}$ stems from the  fact that 
on account of the stretching due to the inflation
all  the co--moving lengths are larger by a factor $\propto R_b(t)$ 
than their non--comoving counterparts.
\end{enumerate}

It is also instructive to give some insight in the qualitative behaviour
of the solution of Equation~\eqref{e-working}.
The coefficients of  Equation~\eqref{e-working} are
singular for $t\to 0$, which is expected because the inflation starts
at $t=0$.
Since $\ddot{R}_b<0$, at early stages the acceleration is dominated by
inflation,  and  Equation~\eqref{e-working} can be approximated by
\be
\label{e-early}
\dd{\phi}{t} + \tfrac{6}{5}\; \frac{1}{t} \d{\phi}{t} +
\tfrac{6}{25} \;\varkappa \; \frac{\phi}{t^2} \sim 0.
\ee
This equation admits analytical solution of the form  $\phi \propto t^\eta$,
where the complex index
\be
\eta = - \tfrac{1}{10} \pm \tfrac{1}{10}\; \sqrt{1 - 24\; \varkappa }.
\ee
is solution to the quadratic equation
\be
\eta^2 + \tfrac{1}{5} \; \eta + \tfrac{6}{25} \; \varkappa  = 0.
\ee
The wavenumber  of a perturbation is clearly limited by the size of
the bubble:  since $\varkappa \geq 1$,
the argument of the square root is always {\em negative}: rearranging the
real and imaginary part of $t^\eta$,
\begin{align}
\label{e-earlydecay}
&\phi(t)  = (t/t_0)^{-1/10} \; \
\left\{
A \cos[\omega\;\log (t/t_0)]
+
B \sin[\omega\;\log (t/t_0)],
\right\}
\intertext{where $t_0$ is  an  arbitrary time  and}
&\omega =  \tfrac{1}{10} \;
\left|24\: \varkappa - 1\right|^{1/2}.
\end{align}
Apart from the oscillating factor,  the co--moving
perturbation amplitude $\phi$
decays as $t^{-1/10}$, showing that the bubble is
Rayleigh--Taylor {\em stable}.  
In non--comoving  coordinates (see Equation~\eqref{e-phidef}) this
the amplitude  grows as $\phi_{\rm phys}\propto t^{1/2}$, i.e. less than 
the bubble radius $R_b\propto t^{3/5}$.
The acceleration is inflation--dominated,
and is directed upwards: this and the relative
stratification of the bubble and the overlying  medium  imply that
the configuration is stable, as found.
Note that the  ``Hubble drag'' term in Equation~\eqref{e-working}
plays a fundamental role in this decay. Without this term
the early--time equation~\eqref{e-early} would read
\be
\label{e-nodrag}
\dd{\phi}{t} +
\tfrac{6}{25} \; \varkappa \; \frac{\phi}{t^2} \sim 0.
\ee
Again,  this admits solutions in the form
\begin{align}
\label{e-nodrag2}
& \phi(t)  = (t/t_0)^{1/2} \; \
\left\{
A \cos[\omega'\;\log (t/t_0)]
+
B \sin[\omega'\;\log (t/t_0)]
\right\},
\intertext{with}
&\omega' =  \tfrac{1}{10} \;
\left| 24 \;\varkappa  -  25 \right|^{1/2}.
\end{align}
In this case (apart from a slowly oscillating term) the
amplitude {\em increases}, albeit at the somewhat slow
pace~$\phi\propto t^{1/2}$. This apparently strange result is
correct, since the energy of the oscillations
diminishes, as the following argument shows.
Equation~\eqref{e-nodrag} is formally the equation of a harmonic
oscillator with a time-dependent  square angular frequency
\be
\label{e-freq}
\Omega^2 = \tfrac{6}{25} \; \frac{\varkappa}{t^2}.
\ee
It is well--known from analytical mechanics (see e.g. \citealp{Arn78}) that
the ratio
\be
I = E/\Omega.
\ee
between the energy $E \sim  \Omega^2\; \phi^2$
of the oscillator
and its angular frequency $\Omega$ is an adiabatic invariant. Therefore,
since $\Omega\propto t^{-1}$, then $\phi\propto t^{1/2}$  and 
\be
\label{e-eng}
E \propto t^{-1},
\ee
which shows that  the energy of the oscillations decreases.

\section{Numerical Results}
\label{s-numerical}

In this Section we solve numerically Equation~\eqref{e-working},
which describes the evolution of the co--moving Lagrangian amplitude $\phi$
(Equation~\eqref{e-phidef})
of a Rayleigh--Taylor perturbation in  a non--inertial 
frame of reference moving  with the expansion of the bubble,
given by Equation~\eqref{e-rtime}.

We refer our numerical calculations to the cluster A~2052: for the
ambient density profile we adopt the $\beta$~-~model
\be
\label{e-dens}
\rho(r) = \frac{\rho_c}{\left[1 + (r/r_c)^2\right]^{3\beta/2}}
\ee
with $\rho_c = 5.7\times 10^{-26}\g \cm^{-3}$, 
$r_c=36.4\kpc $ 
and $\beta=0.56$, which provides an
approximate fit to the deprojected density profile  given by
\citet{Bla01}.

For the  gravitational acceleration  we assume the  \citet{Nav97} profile
calculated by \citet{Zak03} on the same data by \citet{Bla01}:
\be
\label{e-grav}
g(r) = 2 \: g_0 \; (r/r_s)^{-2}\: \left[\log(1 + r/r_s) - \frac{r/r_s}{1+ r/r_s} \right] 
\ee
where $r_s=340\kpc$ and $g_0=2.8\times 10^{-8} \cm\s^{-2}$.

The radius of the bubble (see also Equation~\eqref{e-rtime} and
\citealp{Sok02b} ) is
\be
\label{e-rbubble}
R_b(t) = 7.8 \kpc \; 
\left(\frac{\dot{E}}{10^{44}\ergs }\right)^{1/5} \;
\left(\frac{t}{10^7\yr }\right)^{3/5} \;
\left(\frac{\rho_a}{10^{-25}\g\cc }\right)^{-1/5}.
\ee

In our calculation we must also  take into account the buoyancy of the 
bubble: the drag force limits the bubble rise velocity to 
the uniform terminal velocity
\footnote{This formula assumes that the interior of the bubble is much 
thinner than the surrounding environment.
}
\be
v_b^2 = \tfrac{8}{3} \; \frac{|g|  \; R_b(t)}{C_D} 
\ee
where (following  \citealp{Chu01}) we set the drag coefficient  
$C_D\sim 0.75$. 
This  velocity is slow enough to allow us to simply update at every 
integration step the values of $|g|$ and $\rho_a$ with the values of
the new environment in which the bubble is embedded.
 
The choice of the initial integration time $t_0$ and of the initial 
condition are  not fully trivial, on account of the singularity of
the equation  for $t=0$.
We choose an ``early'' value of $t_0$, meaning that  at this time
the inflation acceleration $\ddot R_b$ is still dominant over gravity.
At this time we fix the value of $\phi$, the ratio between the 
physical (i.e., non--co-moving) amplitude of the perturbation and the
radius of the bubble (see Equation~\eqref{e-phidef}):
\bs
\label{e-iv}
\be
\label{e-iv1}
\phi(t=t_0) = \phi_0.
\ee
Differentiating Equation~\eqref{e-phidef}  we get 
\be
\label{e-iv2}
\dot{\phi}(t=t_0) = \dot{\phi}_0 = 
\phi_0 \left(\frac{1}{t_p}  - \tfrac{3}{5}\frac{1}{t_0}\right);
\ee
\es
the derivative $\dot{\phi}$ measures the  relative magnitude of the 
growth time scales of the perturbation 
$t_p = \phi_{\rm phys}/\dot{\phi}_{\rm phys}$
and $t_b=R_b/\dot{R}_b = 5 t_0/3 $ of the bubble.
We  have integrated numerically Equation~\eqref{e-working},
with a fifth-order adaptive Runge--Kutta numerical algorithm \citep{Pre92}.
We have explored   its solutions for different values of 
the initial conditions~\eqref{e-iv}.
We have stopped (quite arbitrarily)  the integration  at $t_{\rm max}=10^8\yr$
after the inflation starts. This time is few times 
larger than the life times of the radio bubbles  
estimated  by \citet{Bir04}.
Of course, we have ascertained that our result do not
depend on our  choice of   $t_{\rm max}$.

\section{Discussion of the Numerical Results}
\label{s-discussion}

\begin{figure}
\begin{center}
\includegraphics[width=95mm, angle=0]{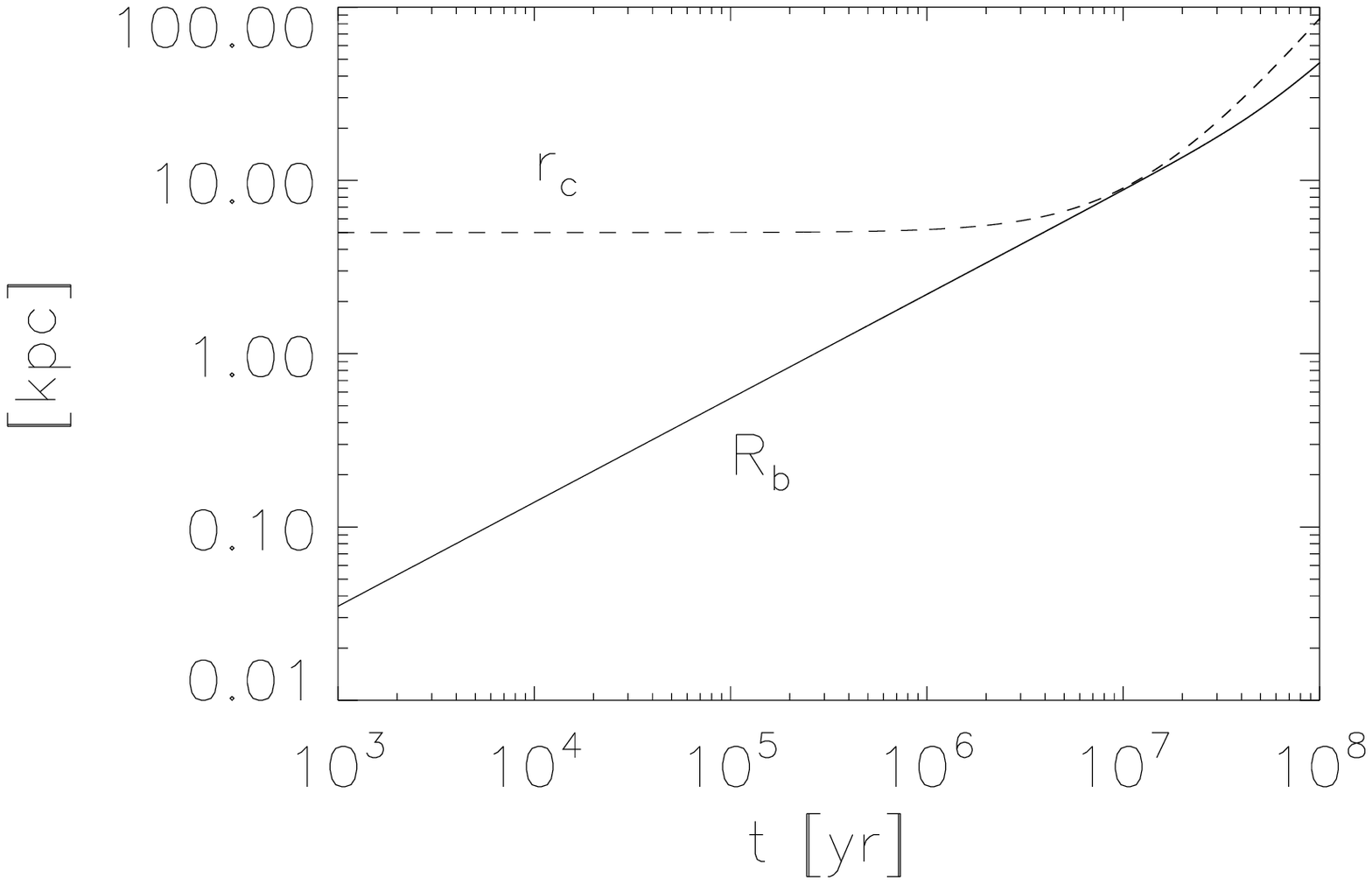}
\includegraphics[width=95mm, angle=0]{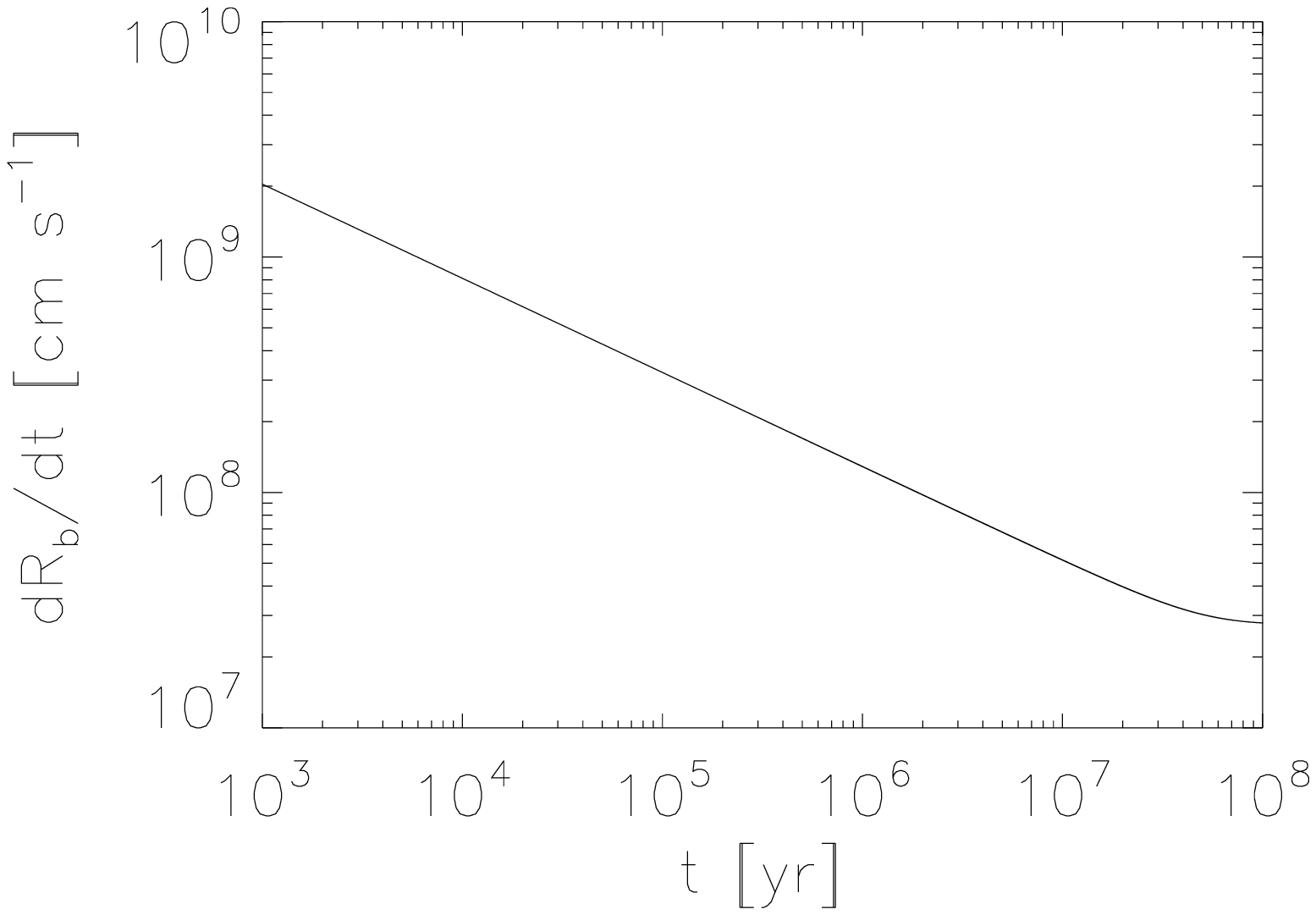}
\includegraphics[width=95mm, angle=0]{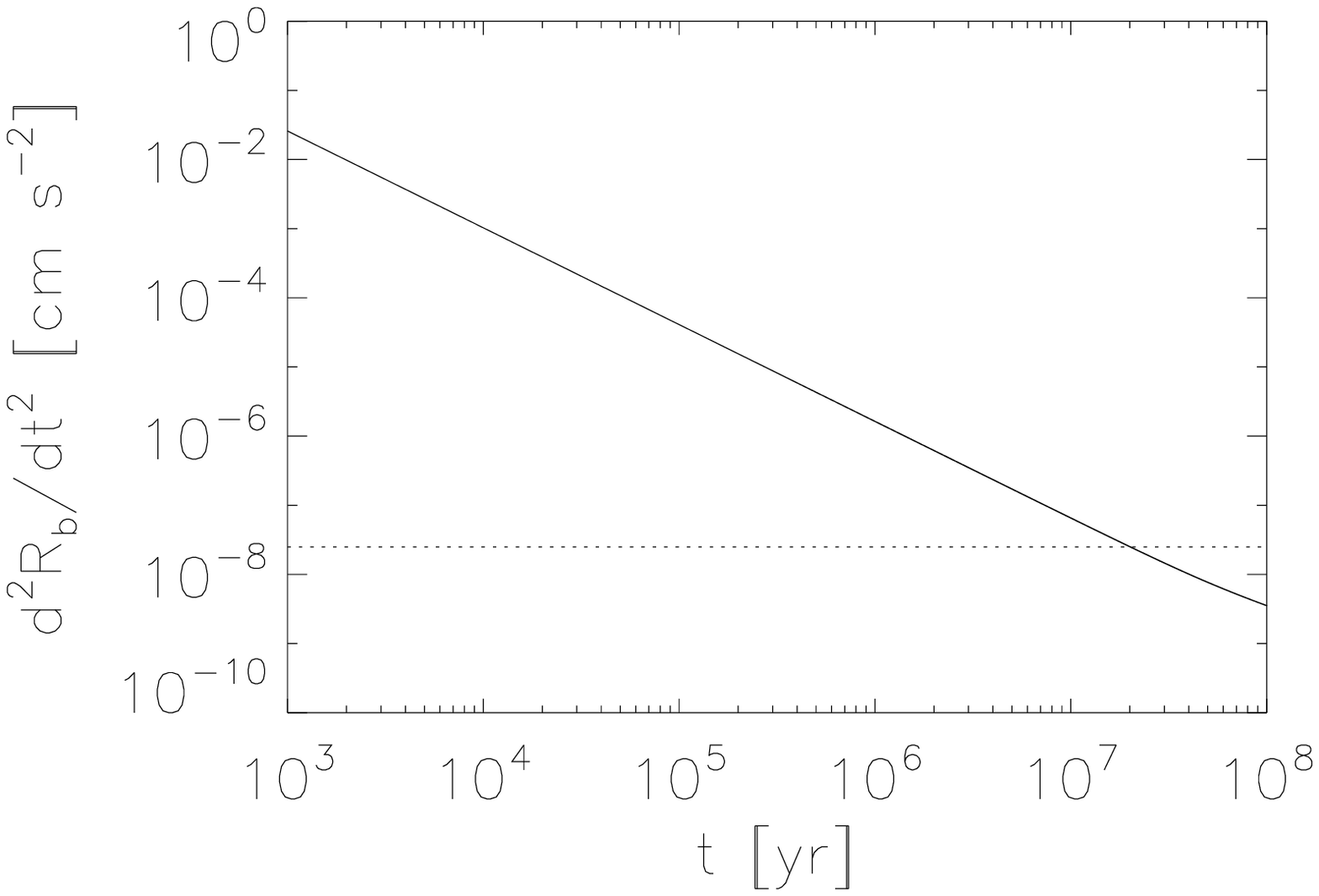}
\end{center}
\caption{\label{f-bubble}
Upper panel: evolution of the radius of a bubble $R_b$ (solid line) compared
to the distance $r_c$ between the centre of the bubble and the centre of the
cluster (dashed line). 
Middle panel: expansion velocity $\dot{R}_b$ of the  bubble front.
Lower panel: acceleration $\ddot{R}_b$ of  the bubble front. 
The dotted  horizontal line shows the modulus $|g|$ of the gravitational 
acceleration at the position where it  equals $|\ddot{R}_b|$
on the top of the bubble.  
In this example the bubble is released at the initial distance $r_c=5\kpc$
from the centre.
}
\end{figure}

\begin{figure}
\begin{center}
\includegraphics{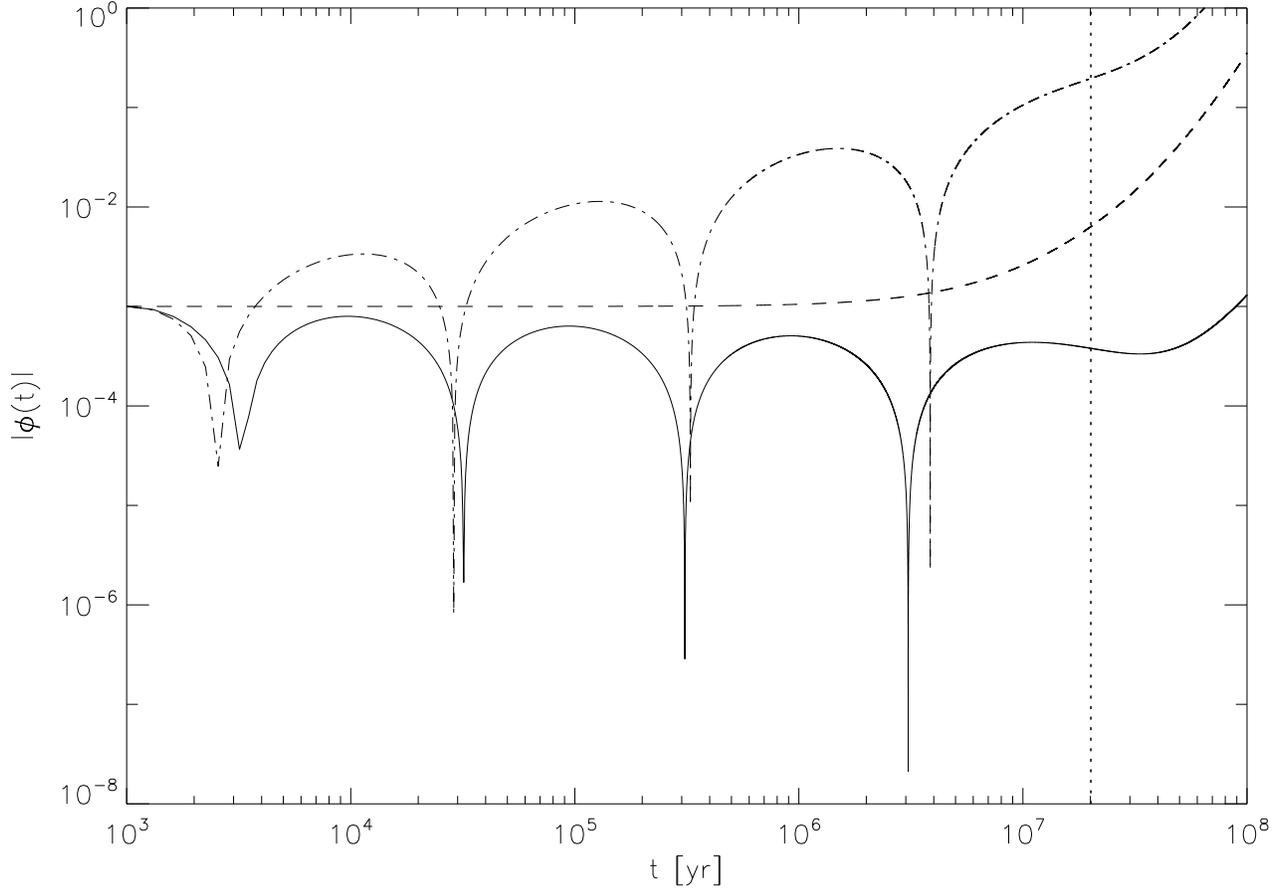}
\end{center}
\caption{\label{f-3curves}
This plot compares the evolution the amplitude of a RTI
of  a perturbation in three cases. Plotted is the absolute value of the 
co--moving perturbation $\phi$, i.e.  the
ratio between the real physical perturbation amplitude and 
the bubble's radius $R_b$ (Equation~\eqref{e-phidef}).
The cusps are  artifacts of the logarithmic scale.
In this example the  bubble is released at $r_c=5\kpc$ from the 
centre, and the ratio between the circumference of the bubble 
and the  wavelength of the physical perturbation is $\varkappa=8$.
The perturbation starts $10^3\yr$ after the bubble is born, with 
an initial co--moving amplitude  $\phi_0=10^{-3}$ and 
zero relative velocity $\dot \phi_0=0$.
The dashed line plots the case in which the inflation's
deceleration is neglected (Equation~\eqref{e-noacc}).
The dot--dashed line plots the evolution of the perturbation without
the ``Hubble drag'' (Equation~\eqref{e-nohubble}). 
The solid line is the evolution described  by the full 
equation~\eqref{e-working}. 
The dotted vertical line marks the time $t_{\rm eq}\sim 2\times 10^7\yr$
when the acceleration 
$\ddot{R}_b$ of the  bubble front
equals in magnitude  the gravitational acceleration 
on the top of the  bubble:
$|\ddot{R}_b|= |g(r_c+R_c)|$.
After~$10^8\yr$ the RT perturbation is well 
developed in the cases described by Equations~\eqref{e-noacc}
and~\eqref{e-nohubble}, but it is still small in the case 
described by Equation~\eqref{e-working}.
}
\end{figure}

\begin{figure}
\begin{center}
\includegraphics{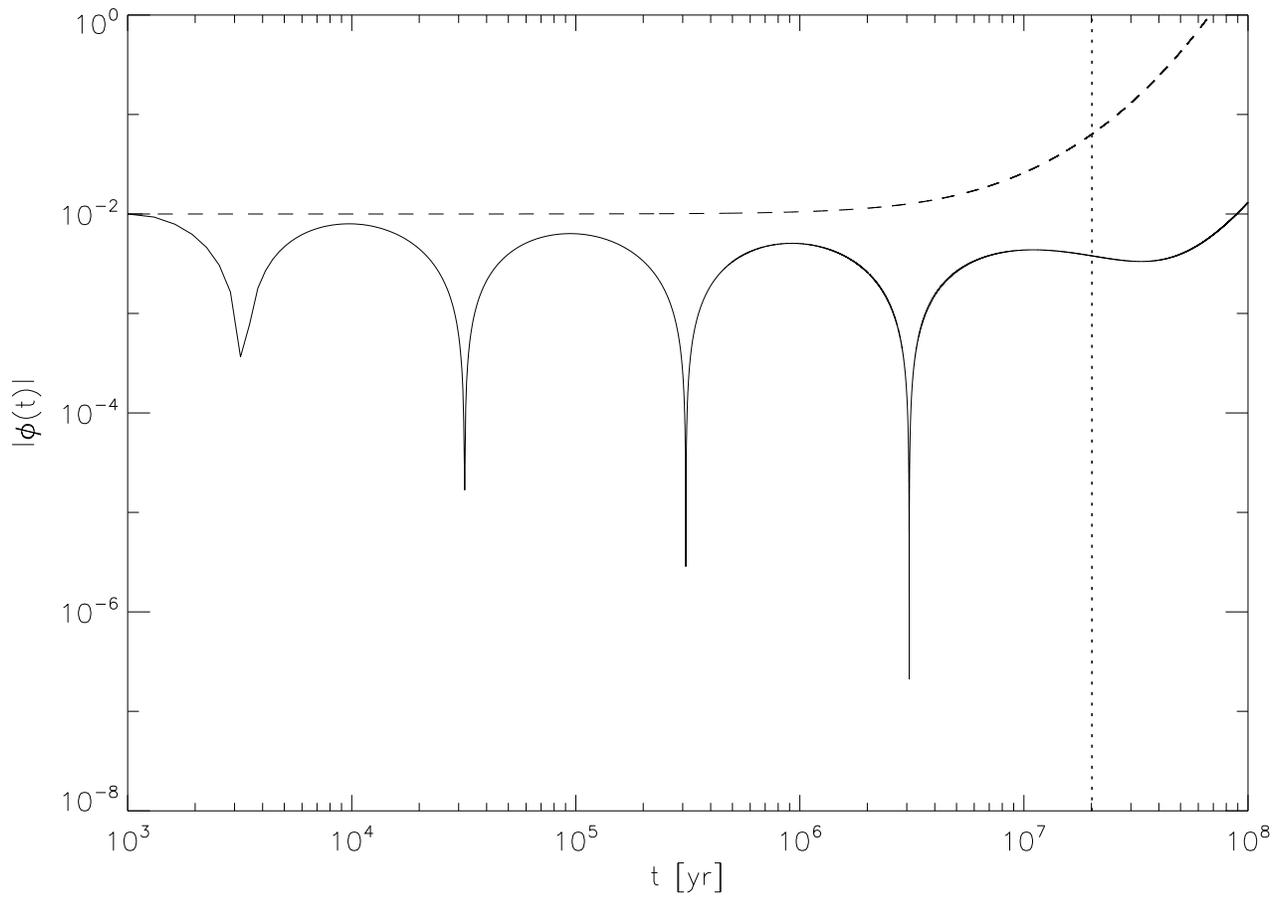}
\end{center}
\caption{\label{f-a0_1E-2} 
The lines have the same meanings as in Figure~\ref{f-3curves};
also the parameters are the same, except the  
the initial relative amplitude, which here  is $\phi_0=10^{-2}$.
}
\end{figure}

\begin{figure}
\begin{center}
\includegraphics{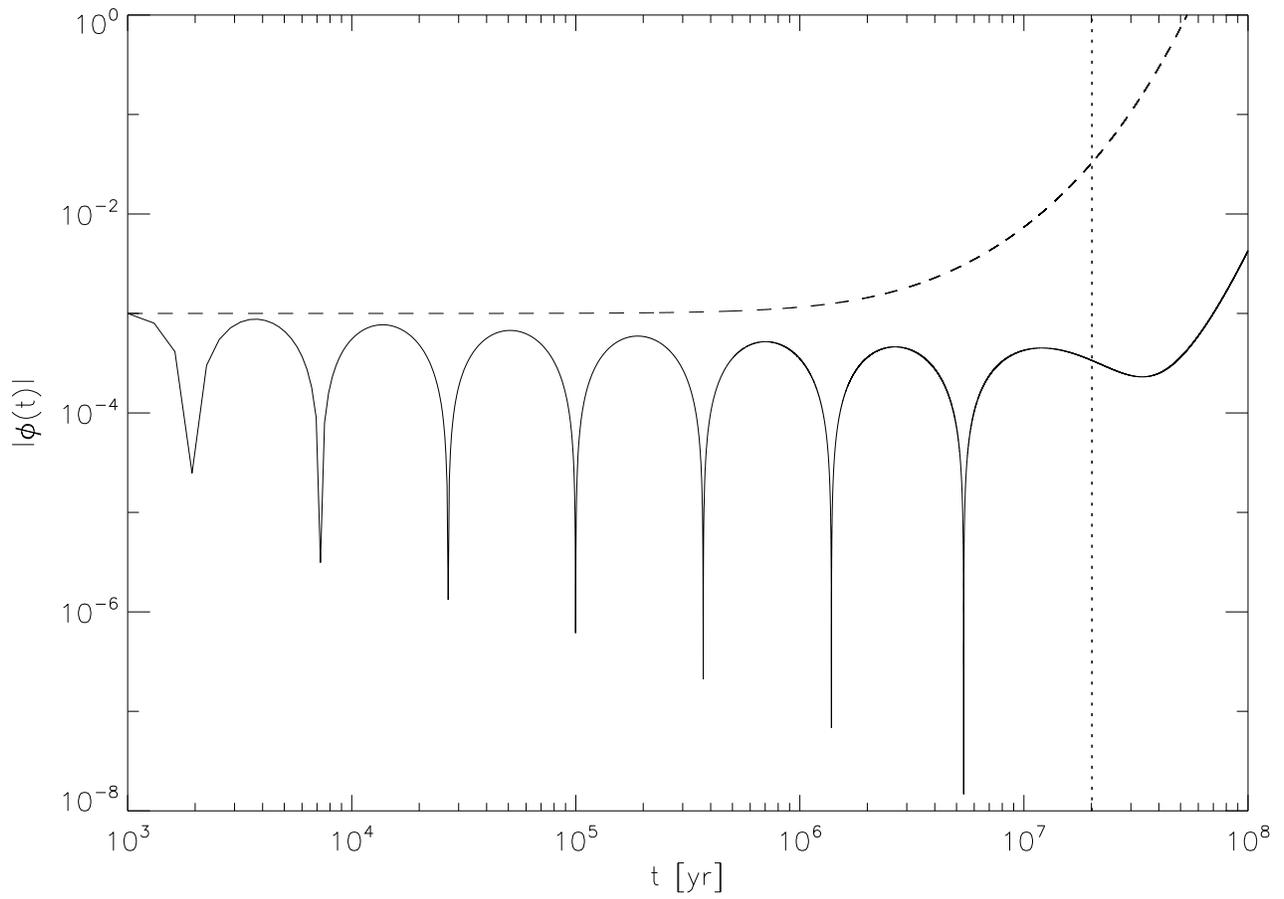}
\end{center}
\caption{\label{f-kR_24.0} 
The lines have the same meanings as in Figure~\ref{f-3curves};
also the parameters are the same, except the  factor
$\varkappa=k_h R_b = 24$. 
The short wavelength perturbations grow faster
(see Equation~\eqref{e-rt}), although they are not responsible for the final
RT disruption of the bubble.
}
\end{figure}

\begin{figure}
\begin{center}
\includegraphics{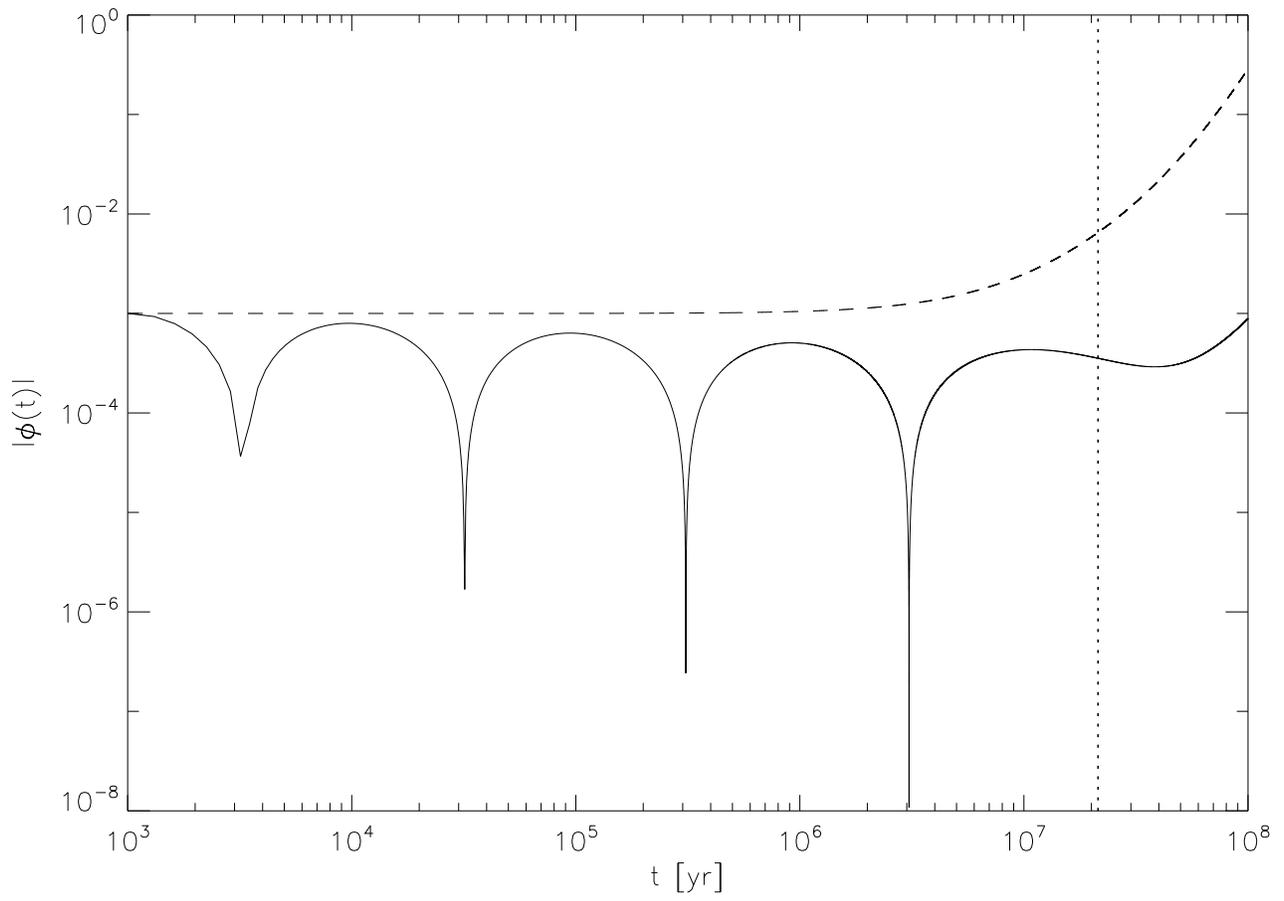}
\end{center}
\caption{\label{f-h0_15.0} 
The same as in Figure~\ref{f-3curves}, but with 
release radius $r_c=15\kpc$.
}
\end{figure}

\begin{figure}
\begin{center}
\includegraphics{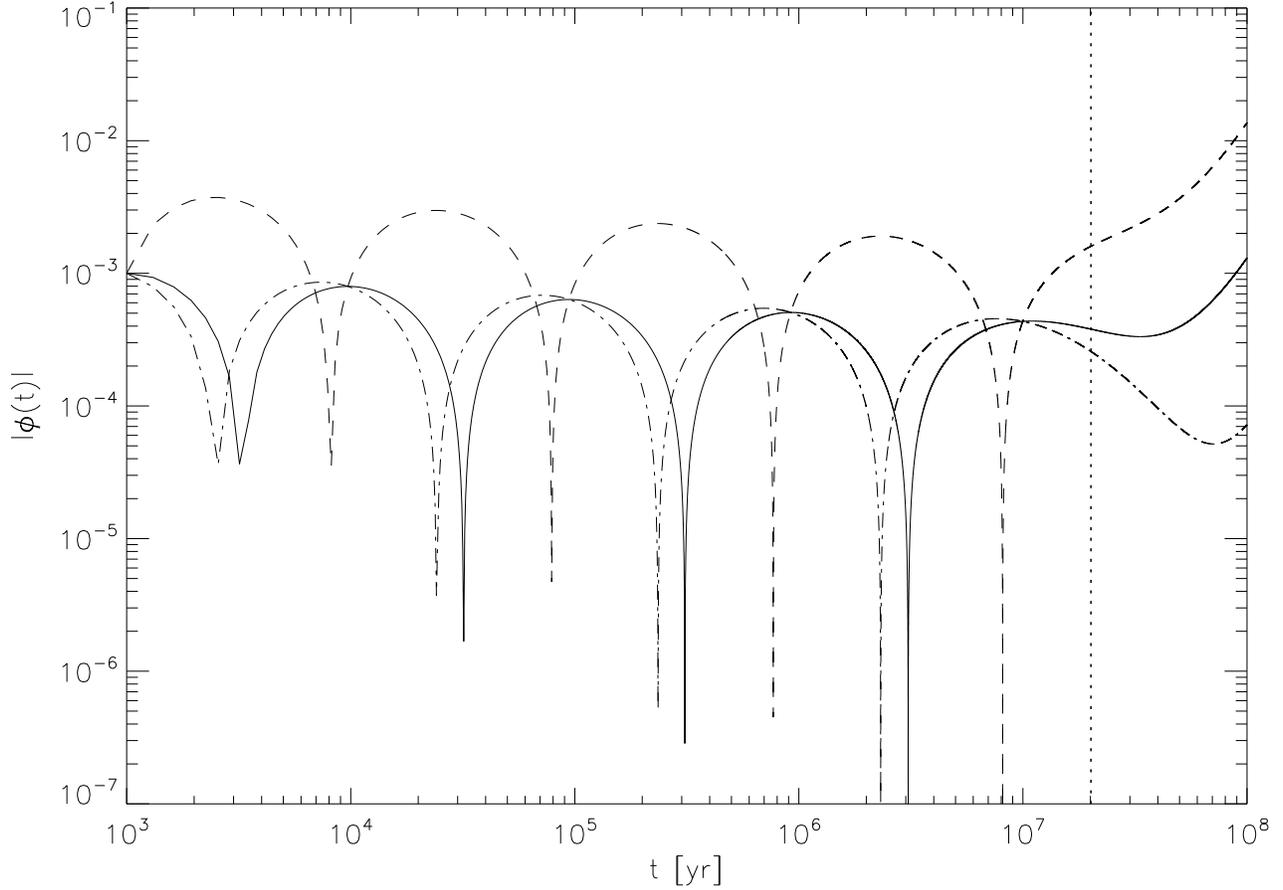}
\end{center}
\caption{\label{f-a1} 
In this graph we show the effect of different initial velocities
on the perturbation's amplitude evolution, as described by
Equation~\eqref{e-working}.
The bubble is released at $r_c=5\kpc$ from the centre.
The integration starts~$10^3\yr$ after the inflation starts, 
with a parameter $\varkappa=8$ and an initial relative 
amplitude~$\phi_0=10^{-3}$. The dashed line corresponds to the
initial condition~\eqref{e-iv2} with $t_p=t_b/10$, the solid line
to $t_p=t_b$, and the dotted--dashed line to $t_p=10\:t_b$.
As expected, the initially fastest--growing perturbations 
develop first, although the effect is not very large.
The perturbations reach the amplitude $|\phi|=0.5$ (at which we may
considered the bubble as disrupted by the instability)
at the times 
$t\sim 3.3\times 10^8\yr$ (case $t_p=t_b/10$),
$t\sim 6.8\times 10^8\yr$ (case $t_p=t_b$),
$t\sim 1.8\times 10^9\yr$ (case $t_p=10\:t_b$);
there is  a factor~$\sim 6$ difference, while the ratio $t_p/t_b$
spans two orders of magnitude.
In all cases, the perturbation develops less than in the case 
without inflation (compare with Figure~\ref{f-3curves}).
}
\end{figure}

\begin{figure}
\begin{center}
\includegraphics{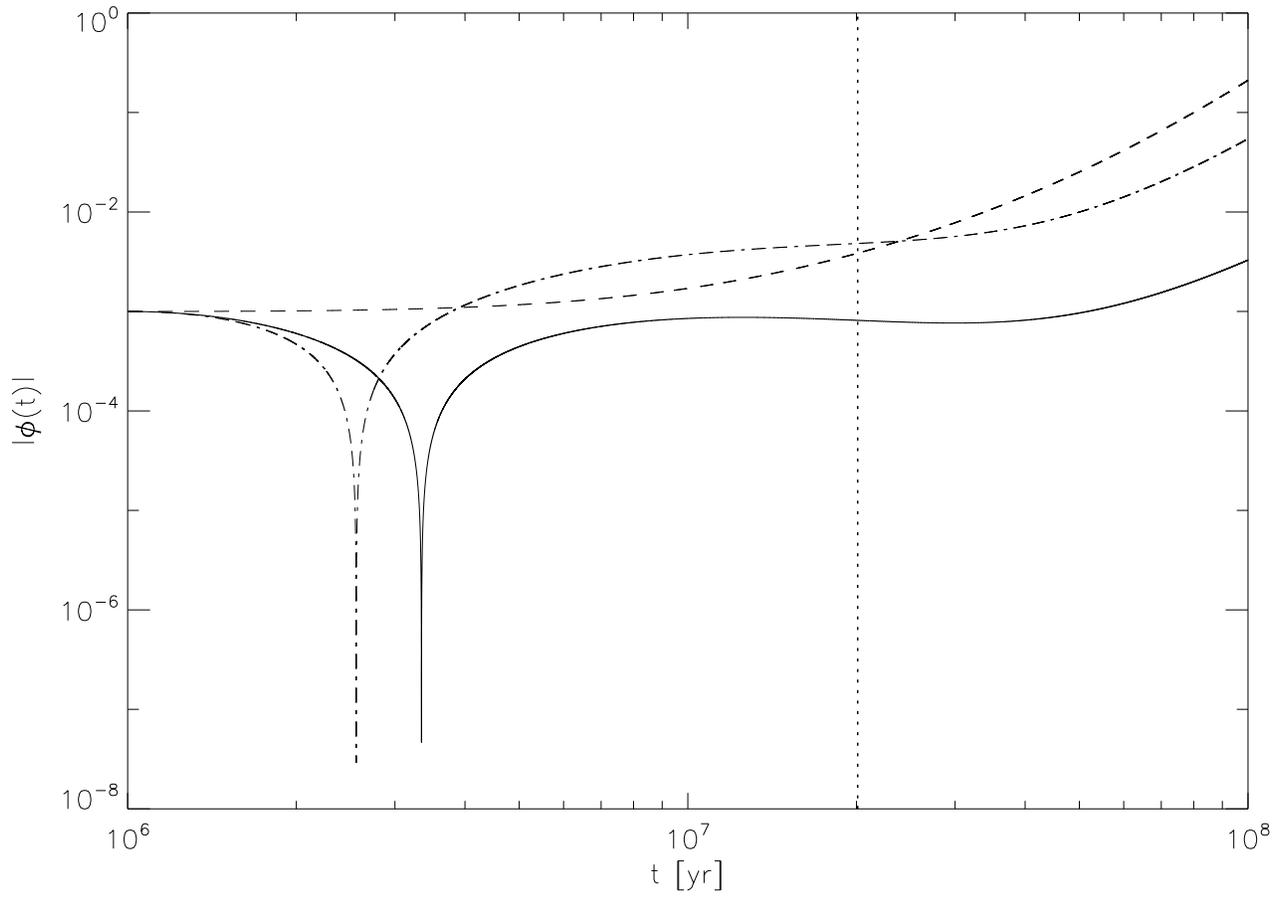}
\end{center}
\caption{\label{f-1e6} 
In this graph we show the effect of the choice of the 
initial integration time
on the perturbation's amplitude evolution, as described by
Equation~\eqref{e-working}.
The parameters and the meaning of the lines are the same  
as in Figure~\ref{f-3curves}, but the 
integration starts $t_0=10^6\yr$ after the inflation onset.
The difference between the case without acceleration
(dashed line, Equation~\eqref{e-noacc}), and the case 
without the Hubble drag 
(dot--dashed line, Equation~\eqref{e-nohubble})
is smaller.
In the case without acceleration the  repulsive force  $\varkappa |g|/R_b(t)$
is smaller, since $R_b$ starts from a larger value.
In the  case without the Hubble drag the oscillations do not
have enough time to develop large enough amplitude before the
time $t_{\rm eq}$ when they become unstable.
}
\end{figure}

\begin{figure}
\begin{center}
\includegraphics{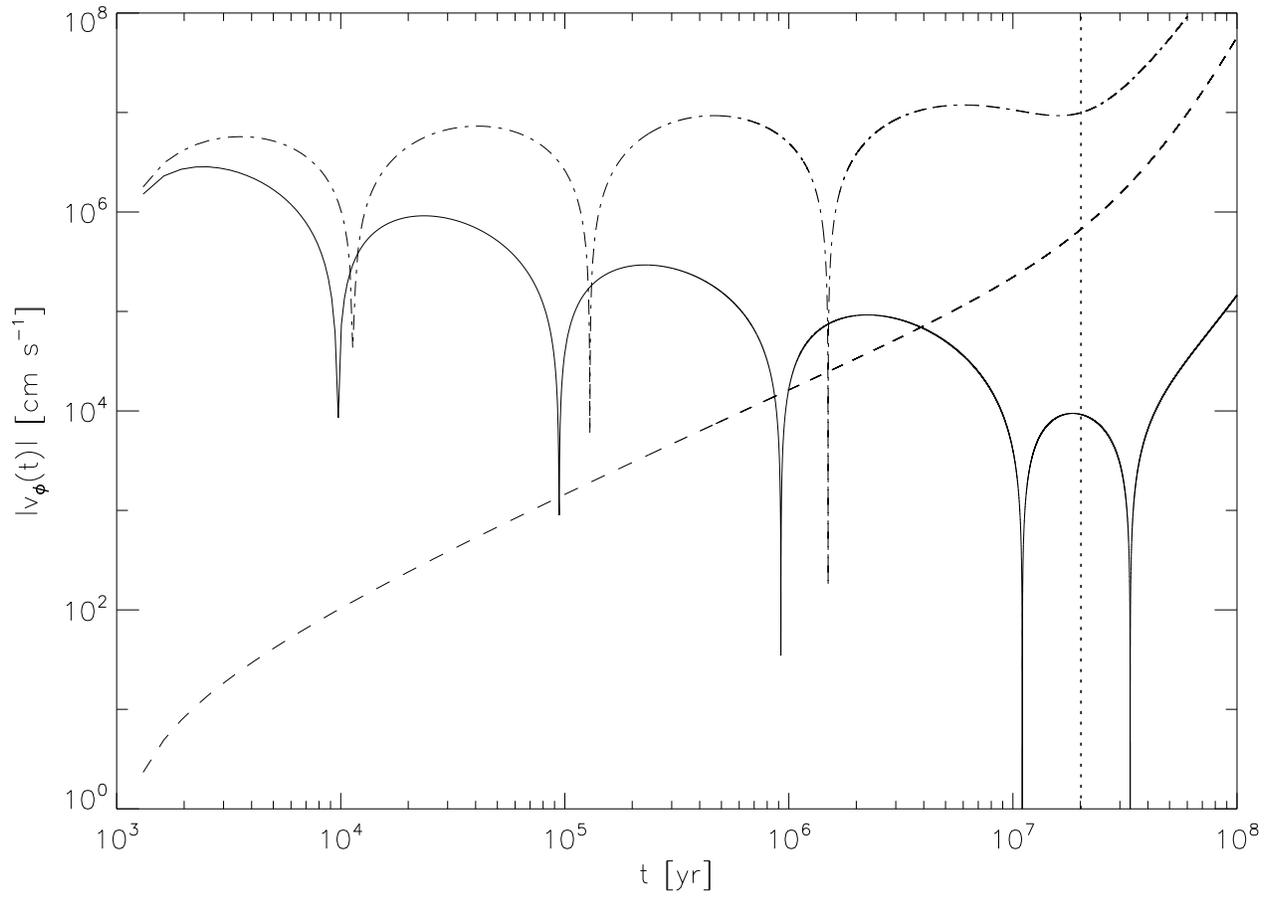}
\end{center}
\caption{\label{f-relvel}
In this graph we show the modulus of the relative velocity 
$v_\phi = \dot{\phi} \: R_b$ of  the perturbation with respect to 
the front of the expanding bubble.  The parameters are the same as
in Figure~\ref{f-3curves}.
}
\end{figure}

In Figure~\ref{f-bubble} we plot the evolution of $R_b$,
$\dot{R}_b$ and $\ddot{R}_b$ for the specific case of a bubble 
released at $r_c=5\kpc$ from the cluster centre.
In Figures~\ref{f-3curves},
\ref{f-a0_1E-2},
\ref{f-kR_24.0},
\ref{f-h0_15.0},
\ref{f-a1}
and
\ref{f-1e6}
we plot the evolution of the co--moving
amplitude $\phi(t)$ for different parameters:
\begin{itemize}
\item
the initial  perturbation amplitude $\phi_0$
(see Equation~\eqref{e-iv1});
\item 
the ratio between the growth time $t_p$ of the perturbation and of the
bubble $t_b$ (see Equation~\eqref{e-iv2};
\item 
the initial time $t_0$ at which the integration starts
\item
the distance $r_c$ from the cluster centre at which the bubble is
initially inflated;
\item
the factor $\varkappa = k_h\:R_b$, i.e. the ratio between the 
circumference of the
bubble and the horizontal wavelength of the  physical RT perturbation.
\end{itemize}
Finally, in Figure~\ref{f-relvel} we have plotted the modulus  of the
relative velocity of the RT perturbation with respect to the
expanding front of the bubble.
In all our calculations we have kept fixed the value of the energy
injection rate $\dot E = 10^{44}\ergs$.

In Figure~\eqref{f-3curves} we show the effect of the different terms
in Equation~\eqref{e-working}.
The dashed line plots the modulus $|\phi|$ of the solution of
Equation~\eqref{e-working} in which the  effects of the
inflationary deceleration are neglected:
\be
\label{e-noacc}
\dd{\phi}{t} -
\varkappa  \; \frac{|g|}{R_b(t)}\;\phi  = 0;
\ee
here inflation is only taken into account as a progressive growth of the 
bubble's radius $R_b$.  
The dot--dashed line plots the solution of Equation~\eqref{e-working}
in which the inflationary deceleration is taken into account, but 
where the ``Hubble drag'' term (i.e., the term proportional to
$\dot\phi$) is neglected:
\be
\label{e-nohubble}
\dd{\phi}{t} -
\varkappa  \; \frac{|g|  + \ddot{R}_b(t)}{R_b(t)}\;\phi  = 0,
\ee
Finally, the solid line plots the solution of Equation~\eqref{e-working}
as it stands.
The ICM density  profile is given by Equation~\eqref{e-dens} and the
gravity profile by Equation~\eqref{e-grav}, and refer to the cluster
A~2052.

In all cases the amplitude grows, leading to the eventual 
bubble   disruption by the RTI. 
At the ``equivalence time'' $t_{\rm eq}\sim 2\times 10^7\yr$ 
when the moduli of the gravitational
acceleration and the inflation deceleration are equal,
the amplitude $|\phi|$ is still small in the case
described by Equation~\eqref{e-working} (solid line), 
but is already well--developed 
in the case described by Equation~\eqref{e-noacc}  (dashed line)
and  in the case described by Equation~\eqref{e-nohubble} 
(dot--dashed line); in this latter case
the perturbation oscillates, and on account of the adiabatic invariance
of the ratio between the oscillations' energy and frequency, the
amplitude increases as $\phi\propto t^{1/2}$
(see the discussion after Equation~\eqref{e-nodrag}).
At $t_{\rm eq}$ the
magnitude of $|\ddot R_b|$ falls below the value of gravity, and
the overall acceleration changes sign;  the RTI sets in and after
few~$t_{\rm eq}$ the bubble is torn apart (but see also 
the next Section for more important details). 

It may seem  a little puzzling that the inclusion of the deceleration
$\ddot R_b$ may  {\em shorten} the life time of the bubble.
This behaviour depends on the magnitude of the quantity
$n^2 = \varkappa \;|g|/R_b$, as we show with the following argument. 
If we ignore the time dependence of $n^2$, then the solution of
\eqref{e-noacc} grows exponentially: 
\be
\label{e-exp}
\phi\propto \eexp{n\:t}.
\ee 
On the other hand, Equation~\eqref{e-nohubble} at small times 
when inflation dominates
has solution 
\be
\label{e-ada}
\phi \propto t^{1/2},
\ee
as discussed in \S~\ref{s-properties}.
In the extreme case $n^2\gg 1$, then the solution~\eqref{e-exp} grows
very fast, while the solution~\eqref{e-ada} is slower.  
At the opposite extreme  $n^2\ll 1$ the solution~\eqref{e-exp}
keeps limited for $t\lesssim n^{-1}$, and is rapidly outgrown by the 
solution~\eqref{e-ada}.
Besides, from  $t_{\rm eq}$ on, the oscillations in Equation~\eqref{e-nohubble}
become unstable. By this time they have developed to a fairly large  
amplitude, which promptly blows up as soon as the instability sets in.
Since the  factor  $n^2 = \varkappa |g|/R_b$ decreases with time 
on account of the bubble's inflation and uprising, the actual situation is
close to the case $n^2\ll1$, which
explains why the system with the bubble's deceleration  included tends
to live  slightly  less than the other.

The solid line plots  the perturbation's evolution as described  by the full 
equation~\eqref{e-working}.
The amplitude oscillates, and initially decays as $\phi\propto t^{-1/10}$ on 
account of the ``Hubble drag'' (discussed after  Equation~\eqref{e-nodrag}). 
After several equivalence times  $t_{\rm eq}$ 
the Rayleigh--Taylor instability finally takes over and  disrupts  the
bubble.
Note that the effect of the Hubble drag is fundamental in  
lengthening  the life time of a bubble. Incidentally,
note that a very similar effect also  occurs in the context of the 
evolution of 
cosmological perturbations; the Hubble  expansion inhibits the 
perturbations, and in particular the density contrast grows
as a mere power--law of time \citep[e.g.][]{Pea99}.

Figure~\ref{f-a0_1E-2} shows the results 
when the 
the initial perturbation is $\phi=10^{-2}$. All other parameters as in 
Figure~\ref{f-3curves}.

In Figure~\ref{f-kR_24.0} the wavelength is one third
that in Figure~\ref{f-3curves};  all other parameters  are the same.
As expected, perturbation growth faster.

In Figure~\ref{f-h0_15.0} the results of 
releasing the bubble at $r_c= 15\kpc$ are  presented.
The bubbles lives a little longer. This will be discussed below.

Changing the initial condition of $\dot \phi$ does not change much the
results, as is shown in Figure~\ref{f-a1}; all other parameters 
as in Figure~\ref{f-3curves}.

Figure~\ref{f-1e6} shows the effect of starting the perturbation at 
$t=10^6\yr$ instead of $t=10^3\yr$; all other parameters as in 
Figure~\ref{f-3curves}.

Finally. in Figure~\ref{f-relvel} we plot the modulus of the relative velocity
$v_\phi = R_b\:\dot \phi$ of the  perturbation with respect to the
bubble front;  note that at early times the perturbation is oscillating: this
is the over-stable regime
for the three cases calculated in Figure~\ref{f-3curves}, which includes 
bubble acceleration, with (solid line) or without (dot--dashed line) 
the drag term.
When the bubble acceleration is
considered without  the drag term (dot--dashed line), the perturbations
attain quite high velocities. In the inner regions of cooling flow clusters
the sound speed is 
\be
c_s = 5.2 \times 10^7  \cm \s^{-1} \; \left(\frac{k\, T}{\keV}\right)^{1/2},
\ee
and therefore the perturbations
oscillate with Mach numbers of $\mathcal M\sim 0.1-0.2$.
At such fast motion some dissipation, which is not included in our 
calculation,
is expected. The dissipation will slow down the growth of the instability in 
the over-stable regime (the oscillatory phase), making the bubble more stable 
already when the drag term is not included (dot--dashed line in 
Figures~\ref{f-3curves} and~\ref{f-relvel}).

From the Figures some robust trends emerge.

\begin{enumerate}

\item {\it Life time.}
The life time of the bubble is always significantly (up to an order of 
magnitude)
longer when the initial inflationary phase is accounted for
\citep{Sok02b}.
For typical values used here, the bubbles are torn apart by the RTI after
a time $\ga 10^8 \yr$, assuming that the jet is active for 
$\sim 3\times 10^7-10^8\yr$ \citep{Bir04}.
This is enough to account for the existence of outer  ghost bubbles
in clusters of galaxies, whose ages are $\la 10^8 \yr$ 
\citep{Dun05}.

\item {\it Bubble's location. }
The bubble injected close to the cluster centre live shorter than those 
injected at larger distances. The reason is that the growth time of the 
RTI is  proportional to $\lambda_h^{1/2}$ (Equation~\eqref{e-rt}). 
Bubbles near the centre expand at a lower rate
(see Equation~\eqref{e-rtime}), hence for a constant ratio  
$k_h R_b=2 \pi R_b /\lambda_h$ bubbles  near the centre
have shorter perturbation wavelength $\lambda_h$, and therefore the RTI 
evolves faster.
Also, near the centre gravity is higher, also shortening the RTI growth 
time. Note that the  effect of gravity is not very large in  the
\citet{Nav97} gravitational potential
we have considered, but  would be more relevant in presence of 
a steeper gravitational profile.

\item {\it Wavelength. }
The instabilities  with short wavelength (i.e. higher value of 
$\varkappa=k_h\:R_b$)
develop first (see also Equation~\eqref{e-rt}).
They may change
somewhat the bubble boundary, but it is the large wavelength
perturbations, $\lambda_h \sim R_b$ that will tear  the bubble apart.

\item{\it Energy Injection Rate}
We have run some calculations with a lower energy injection rate $\dot E$,
without reporting the relative figures, since they are very similar to those
already presented. As one  might expect, a lower energy injection rate
entails a weaker deceleration and a weaker drag, and hence a shorter
bubble lifetime. For $\dot E=10^{42}\ergs$, for instance (i.e. two
orders of magnitude less than the case presented in the paper), 
the equivalence time only halves: $t_{\rm eq}\sim  10^7\yr$ instead of
$t_{\rm eq}\sim 2\times 10^7\yr$. The difference is not large, and is due to
the weak dependence of the bubble radius on the energy injection rate
($R_b\propto {\dot E}^{1/5}$, see Equation~\eqref{e-rtime}).

\end{enumerate}

Before ending this Section, it is worth to compare our results with the
numerical simulations of \citet{Bru02}, who studied numerically the 
evolution of radio bubbles during the active inflation phase. 
A detailed comparison is difficult, since  \citet{Bru02}
include the effects of a possible Kelvin--Helmholtz instability (KHI), 
which we have neglected.  The KHI resulting from the bubble shear velocity 
with respect to  the outer environment shapes the bubble as a mushroom,
especially for low energy powering AGN jets. The RTI, on the other hand,
tends to disrupt the top of the bubble first.
As it is apparent from Figures~2, 4 and~5 of \citet{Bru02}, the top of the
bubble keeps its round shape even if the jet is weak, showing that the
RTI is not very relevant at this stage.
Also the dependence of the lifetime of a bubble  on the $\dot E$
supplied by the jet agrees qualitatively with our results, although 
a more quantitative assessment is difficult.
We  also note that the flow of the ICM along the bubble sides as 
the bubble rises, will stretch the magnetic field lines along the flow 
direction. This builds the right tension to suppress the KHI
\citep{DeY03}. 
It is much more difficult to build a magnetic field to stabilise 
the top front against the RTI, however. 
Therefore, although we claim than magnetic 
fields are not required to protect  the bubble against the RTI, we do 
accept that they  can suppress the KHI.

\section{The Post--Inflation Phase}
\label{s-last}

The results presented in the previous Sections refer to an early phase of
the lifetime of a bubble, when it is still powered by the jet
launched by the central AGN. It is important to stress that this is a 
crucial assumption for our model. Yet, the jet cannot power the
expansion indefinitely, and after  few $10^7\yr$ 
\citep[e.g][]{Ale87, Bir04} the jet fades.
At the end of this  ``active''  inflation phase 
the  amplitude of the perturbations  has been slightly suppressed
by the drag and deceleration effects
(see Equation~\eqref{e-earlydecay}, or Figure~\ref{f-3curves}),
so  the instabilities will take more time to develop fully.  
After the jet switches off,  the bubble rises buoyantly due to its low density.
The process is almost adiabatic (there are but negligible heat exchanges
between the bubble and the environment), so the bubble's  expansion is
far too slow  to induce a deceleration or a drag 
force strong enough to hinder the growth of the RTI. 
The RTI sets in, leading to the eventual disruption of the bubble itself.

At this stage the evolution of the bubble is quite difficult to 
follow analytically, on account of  a host of new dynamical effects
which will be discussed below. We refer to the existing literature
for the numerical simulations of the evolution of these
buoyant bubble, no more powered by an active jet
(we refer in particular to  \citealp{Bru01} and \citealp{Rey05}: 
since they  do not include magnetic fields 
in their simulations, their results are more directly comparable to the
case analysed in the present work).

In this Section we  address briefly the most relevant physical effects
occurring in the buoyant phase, as well as 
their  influence on  the growth of the bubbles' instabilities.

After the end of the inflation's stabilising effect (or after few
times $t_{\rm eq}$, when the deceleration and the drag are too weak
to be relevant)
the RTI is free to evolve. As the  perturbations' amplitude
attains the magnitude
$\phi_{\rm phys}/\lambda_h\sim 0.1-0.2$, i.e. $\phi\sim 0.5-1  / \varkappa$ 
the RTI enters the non--linear regime.
\citet{Ofe92} have shown that  if the RT perturbations are not exactly 
monochromatic (which is the case in a realistic situation) the coupling 
between modes of different wavelengths suppresses the 
fast--growing short--wavelength modes; the 
overall effect is to  suppress the growth of the instabilities,
resulting in a further delay in  the bubble disruption.

During the non--linear evolution  the relative velocity
of the bubble with respect to the ambient ICM may significantly 
affect the dynamics, with potentially dominant  
contributions from the KHI.
There is another reason why the KHI becomes important only at this stage:
in the early inflation phase the  density contrast  between the bubble 
interior and the ambient gas  is very large, and in this situation the 
KHI is suppressed \citep{Kai05}. Later on, however, when the density 
contrast lowers,
the KHI  may become important enough to be the primary cause for the 
bubble disruption.

Some authors suggest that the KHI growth rate may be reduced or 
suppressed in this phase.  \citet{Kai05} and \citet{Rey05} show
that the transport phenomena (plasma viscosity and thermal conduction)
are able to suppress the growth of the fastest Rayleigh--Taylor and
Kelvin--Helmholtz modes. This result is particularly important in the
context of the present paper, since in the absence of an ordered
magnetic field  parallel to the bubble surface  neither viscosity nor
thermal conduction are expected to be strongly suppressed.

In the  different context of the stability of cold fronts against the KHI
\citet{Chu04} have shown that  if the interface between the interior of 
the bubble and the ICM has a finite thickness, the fast  
short--wavelength  KHI modes  are suppressed.
As mentioned before, we do not expect the magnetic field to suppress the
RTI, but we do expect it is able to reduce the growth of the KHI  modes.

To summarise, even in the buoyant phase the growth of the instabilities 
may be reduced by the occurrence of several effects. We repeat  that 
we do not expect them to prevent the final disruption of the bubble.
The bubble {\em will} be torn apart, but on time scales of few $10^8\yr$,
consistent with the estimated ages of the ``ghost'' bubbles observed in 
several cooling flow clusters. 
The final onset of the RTI, with consequent disruption of the bubble,
is only delayed but not avoided, which
is consistent with ghost bubbles showing RTI features. 
\citet{Sok02b} suggest that the protrusion in the northwest bubble of
the Perseus cluster \citep{Fab00, Fab02}  is a clear signature of
the   late onset of the Rayleigh--Taylor instability.

\section{Summary}
\label{s-summary}

To summarise, our main conclusion is that the life time for 
X--ray deficient bubbles in
cluster of galaxies, and the development of RTI features in ghost bubbles
can be explained from pure gas dynamical effects, and there is no
need to assume the existence of stabilising magnetic field.
This conclusion is robust, and is not affected by the several simplifying
assumptions we have made.
For typical parameters, inclusion of the bubble inflation phase increases
the expected life of bubbles by a factor of~$5-10$.

\section*{Acknowledgements}
It is a pleasure to thank Adi Nusser for his fruitful observations.
We also thank an anonymous referee, whose  suggestions and criticism
helped us to improve this paper.
This research was supported in part by the Asher foundation at the 
Technion.


\newpage



\newpage

\bsp

\label{lastpage}

\end{document}